%Paper: hep-th/9211073
%From: gcleaver@theory3.caltech.edu (Gerald Cleaver)
%Date: Mon, 16 Nov 92 13:32:15 PST

%%%%%%%%%%%%%%%%%%%%%%%%%%%%%%%%%%%%%%%%%%%%%%%%%%%%%%%%%%%%%%%%%%%%%%%%%%%
%  Gerald B. Cleaver and David C. Lewellen
%  ``On Modular Invariant Partition Functions for Tensor Products of
%    Conformal Field Theories.''
% _________________________________________________________________________
% Some macros for HEP preprints
\input phyzzx
\hoffset=1truein
\voffset=1.0truein
\hsize=6truein
\def\TITLEPAGE{\frontpagetrue}
\def\CALT#1{\hbox to\hsize{\tenpoint \baselineskip=12pt
	\hfil\vtop{\hbox{\strut NSF--ITP--92-148}
        \hbox{\strut CALT--68-#1}
	\hbox{\strut DOE RESEARCH AND}
	\hbox{\strut DEVELOPMENT REPORT}}}}

\def\ABSTRACT#1{\vskip .5in \vfil \centerline{\twelvepoint \bf Abstract}
	#1 \vfil}
\def\ENDTITLEPAGE{\vfil\eject\pageno=1}

\def\sqr#1#2{{\vcenter{\hrule height.#2pt
      \hbox{\vrule width.#2pt height#1pt \kern#1pt
        \vrule width.#2pt}
      \hrule height.#2pt}}}

\def\section#1#2{
\noindent\hbox{\hbox{\bf #1}\hskip 10pt\vtop{\hsize=5in
\baselineskip=12pt \noindent \bf #2 \hfil}\hfil}
\medskip}
%___________________________________________________________________________

\voffset 0.8truecm\hoffset 0.8truecm
\def\Z{Z\!\!\!Z}

\def\hbsev #1{\hbox to .7cm{#1}}
\TITLEPAGE
\CALT{1754}
\vskip 1.2truecm
\titlestyle {On Modular Invariant Partition Functions for
\break
Tensor Products of Conformal Field Theories\foot{Work supported
in part by the U.S. Dept. of Energy
under Contract no. DEAC-03-81ER40050 and the National Science Foundation under
grant no. PHY 89-04035.}}
\vskip .5truecm
\centerline{Gerald B. Cleaver}
\centerline{\it California Institute of Technology, Pasadena, CA 91125}
\centerline{\it and}
\centerline{David C. Lewellen}
\centerline{\it Institute for Theoretical Physics, Santa Barbara, CA 93106}
\ABSTRACT{We give two results concerning the construction of modular invariant
partition functions for conformal field theories constructed by tensoring
together other conformal field theories. First we show how the possible modular
invariants for the tensor product theory are constrained if the allowed modular
invariants of the individual conformal field theory factors have been
classified. We illustrate the use of these constraints for theories of the type
$SU(2)_{K_A}\otimes SU(2)_{K_B}$, finding all consistent theories for $K_A,K_B$
odd. Second we show how known diagonal
modular invariants can be used to construct some inherently asymmetric ones
where the holomorphic and anti-holomorphic theories do not share the
same chiral algebra. Some explicit examples are given. }

\ENDTITLEPAGE

\chapter{Introduction}

    In the past few years considerable effort has been spent in classifying
modular invariant partition functions of two-dimensional conformal field
theories.  Complete classifications exist only for some of the simplest
conformal field theories, in particular the minimal discrete series with $c<1$,
and the
models based on level $K$ SU(2) Ka\v c-Moody algebras.$^{[1]}$
  For string theory
applications, conformal field theories with larger central charges (up to 26)
are of more direct interest, so it is natural to consider the possibilities for
constructing such theories using tensor products of the simpler, well
known theories. To date this program has been systematically carried out only
for theories constructed from free bosons or fermions.$^{[2,3]}$
For tensor products of
other theories no procedures have been developed which give all of the possible
modular invariants, but a few simple algorithms exist for modifying a known
modular invariant to produce another one, in particular the orbifold
construction$^{[4]}$ and the related
operation of twisting by a simple current.$^{[5]}$

    In this work we make some modest proposals aimed at the general problem of
classifying all possible modular invariants for conformal field theories
constructed by tensoring together models whose modular invariants are already
known. By a tensor product of two theories, say $A$ and $B$, we mean a theory
whose chiral algebra includes the chiral algebras of both the $A$ and $B$
theories. As a consequence, the central charge of the combined theory will be
the sum of those for the individual factors, the chiral blocks which make up
amplitudes will be constructed from the products of the individual chiral
blocks, and the characters will be
products of the individual characters, restricting the partition function to
the form,
$$Z^{AB}=\sum_{l,m,\bar{l},\bar{m}}
N^{AB}_{lm\bar{l}\bar{m}}\chi^A_l\chi^B_m\bar{\chi}^A_{\bar{l}}
\bar{\chi}^B_{\bar{m}}\quad .\eqn\zab$$

The combined theory is {\it not} restricted to be simply the product of the
individual theories; the operators in the combined theory need not
be diagonal (i.e. left-right symmetric), and in general the fusion rules for
the operator products will be modified. The latter point is the chief
complication in the general problem. The allowed tensor product theories built
from free bosons or fermions have been successfully categorized because the
possible fusion rules in these theories are almost trivial; likewise
twisting a theory by a simple current gives unambiguously a new theory because
the new fusion rules are unambiguous.

    In section 2 we consider to what extent the integer coefficients
$N^{AB}_{lm\bar{l}\bar{m}}$ in the partition function of the tensor product
theory are constrained if we know all of the allowed possibilities for the
corresponding coefficients $N^A_{l\bar{l}}$ and $N^B_{m\bar{m}}$ in the factor
theories. In section 3 we consider the more general possibility of combining
theories such that the holomorphic and anti-holomorphic degrees of freedom need
not possess the same chiral algebras, that is we consider partition functions
of the form,
$Z=\sum_{l,\bar{m}}N_{l\bar{m}}\chi^A_l\bar{\chi}^B_{\bar{m}}$.
In the following
sections we are interested ultimately
in classifying consistent conformal field theories, not just modular invariant
combinations of characters. Accordingly, we freely invoke consistency
conditions for amplitudes on the plane when they prove useful for constraining
the states which can appear in the partition function.

\chapter{Constraints on Tensor Product Modular Invariants}

   In order for the tensor product partition function \zab\ to be invariant
under the generators of modular transformations $\tau\rightarrow\tau+1$ and
$\tau\rightarrow-1/\tau$ (denoted $T$ and $S$, respectively) we must have,
$$\eqalign{T\quad{\rm invariance:}\quad &\Delta_l+\Delta_m=\Delta_{\bar{l}}+
\Delta_{\bar{m}}\quad ({\rm mod}\quad 1)\quad {\rm if}\quad
N^{AB}_{lm\bar{l}\bar{m}}\ne 0\cr
S\quad{\rm invariance:}\quad &N^{AB}_{lm\bar{l}\bar{m}}=\sum_{l^\prime,
m^\prime,\bar{l}^\prime,\bar{m}^\prime} N^{AB}_{l^\prime m^\prime
\bar{l^\prime}\bar{m^\prime}}S^A_{ll^\prime}S^B_{mm^\prime}
\bar{S}^A_{\bar{l}\bar{l}^\prime}\bar{S}^B_{\bar{m}\bar{m}^\prime}
\quad .\cr}\eqn\minv$$
   We assume that the solutions to the corresponding equations for the
factor theories are already known. That is, we know all possibilities
(labeled by
$i$) for non-negative integer coefficients $N^{A,i}_{l\bar{l}}$ such that,
$$\eqalign{&\Delta_l=\Delta_{\bar{l}}
\quad ({\rm mod}\quad 1)\quad {\rm if}\quad N^{A,i}_{l\bar{l}}\ne 0\cr
{\rm and}\quad\quad&N^{A,i}_{l\bar{l}}=\sum_{l^\prime,
\bar{l}^\prime} N^{A,i}_{l^\prime  \bar{l^\prime}}S^A_{ll^\prime}
\bar{S}^A_{\bar{l}\bar{l}^\prime}\quad ,\cr}\eqn\aminv$$
and similarly for $N^{B,j}_{m\bar{m}}$.  We can get relations between the
integer coefficients in equations \minv\ and \aminv\
by multiplying \minv\ by $N^{A,i}_{l\bar{l}}$ and summing over $l$ and
$\bar{l}$,
$$\eqalign{\sum_{l,\bar{l}}N^{A,i}_{l\bar{l}}N^{AB}_{lm\bar{l}\bar{m}}
&=\sum_{l,\bar{l},l^\prime,
m^\prime,\bar{l}^\prime,\bar{m}^\prime}N^{A,i}_{l\bar{l}}
 N^{AB}_{l^\prime m^\prime
\bar{l^\prime}\bar{m^\prime}}S^A_{ll^\prime}S^B_{mm^\prime}
\bar{S}^A_{\bar{l}\bar{l}^\prime}\bar{S}^B_{\bar{m}\bar{m}^\prime}\cr
&=\sum_{l^\prime,
m^\prime,\bar{l}^\prime,\bar{m}^\prime}N^{A,i}_{l^\prime\bar{l^\prime}}
 N^{AB}_{l^\prime m^\prime
\bar{l^\prime}\bar{m^\prime}}S^B_{mm^\prime}
\bar{S}^B_{\bar{m}\bar{m}^\prime}
\quad ,\cr  }\eqn\amom$$
where we have used \aminv\ and the symmetry of $S$ to simplify the right
hand side. The resulting
equation is precisely of the form \aminv\ for the $B$ theory, therefore we must
have,
$$\sum_{l,\bar{l}}N^{A,i}_{l\bar{l}}N^{AB}_{lm\bar{l}\bar{m}}
  =\sum_jn^{A,i}_j N^{B,j}_{m\bar{m}}\quad .\eqn\constr$$
This constrains some combinations of coefficients in the $AB$ theory to be
linear combinations (with integer coefficients) of the  allowed
coefficients in the $B$ theory which are presumed known. There is an analogous
constraint arising from taking the appropriate traces over the $B$ theory
indices in \minv, and a further constraint arising from taking appropriate
traces in both sets of indices in either possible order. Note that the number
of constraint equations increases as the factor theories become more complex
(in the sense of having more possible modular invariants), and also as we
consider tensor product theories with more factors.

    These equations constrain part of the
operator content of the tensor product theories which we wish to classify.
Often, this information, together with some simple consistency requirements for
conveniently chosen amplitudes on the plane, serves to completely determine the
allowed possibilities for the tensor product modular invariants. For
concreteness, we illustrate with a simple example.

   Example: SU(2)$_{K_A}\otimes$SU(2)$_{K_B}$ tensor product theories.

 There are $K+1$ unitary primary fields of  $SU(2)_{K}$, which we will label by
twice the spin, $l=2s$, of the corresponding SU(2) representation.
The conformal dimensions are $\Delta_l= {{l(l+2)}\over 4(K+2)}$.
The matrix $S$ implementing the modular transformation $\tau\rightarrow-1/\tau$
on the Ka\v c-Moody characters is,
$$S^K_{ll^\prime}=\left(2\over K+2\right)^{1/2}\sin\left(\pi (l+1)(l^\prime +1)
\over K+2\right)\eqn\ssu$$ and the fusion rules, which we will make use of
momentarily are,
$$\phi_l\times\phi_{l^\prime}=\sum^{{\rm min}(l+l^\prime,2K-l-l^\prime)}_{
{m=\vert l-l^\prime\vert \atop m-\vert l-l^\prime\vert\,\,{\rm even}}}
\phi_m \quad .\eqn\fra$$

   For simplicity we will only consider the possibilities for tensor product
theories with holomorphic and anti-holomorphic chiral algebras
SU(2)$_{K_A}\otimes$SU(2)$_{K_B}$ for both $K_A$ and $K_B$ odd. Then the only
possible modular invariants for the factor theories are the diagonal ones,
$N_{l\bar{l}}=\delta_{l\bar{l}}$.\foot{For our purposes we need all
non-negative integer coefficients $N_{l\bar{l}}$ which give rise to $S$ and
$T$ invariant partition functions, but not necessarily with a unique vacuum
state ($N_{00}=1$). We have confirmed in the SU(2) case that relaxing this
condition does not expand the space of possible solutions beyond a
multiplicative constant.} Applying the constraint equation \constr\ and its
obvious generalizations gives the conditions,
$$\eqalign{\sum_{l=\bar{l}}
N^{AB}_{lm\bar{l}\bar{m}}&=a\delta_{m\bar{m}}\,\, ;\quad a\in \Z^+\cr
 \sum_{m=\bar{m}} N^{AB}_{lm\bar{l}\bar{m}}&=b\delta_{l\bar{l}}\,\, ;\quad\quad
b\in \Z^+\cr
\sum_{l=\bar{l}}\sum_{m=\bar{m}} N^{AB}_{lm\bar{l}\bar{m}}&=a(K_B+1)=b(K_A+1)
\cr}\eqn\nabcon$$

    If we label the primary operators in the tensor product theory by the
corresponding $l$ values of the factor theories, e.g.
$(l,m\vert\bar{l},\bar{m})$, then the integer $a$ is equal, in particular, to
the number of primary operators in the theory of the form $(j,0\vert j,0)$.
These are pure $A$ theory operators and so must form a closed operator
subalgebra of the $A$ theory. Similarly, $b$ must equal the dimension of some
closed operator algebra in the $B$ theory. This is useful because we know
(explicitly from studying the consistency of amplitudes on the plane) all
consistent closed operator sub-algebras in
SU(2) Ka\v c-Moody theories.$^{[6]}$  For $K$
odd these are (labeled by their dimensions, $d$),
$$      \eqalign{d = 1: &~~\{\Phi_0\} {\rm \, \, (the~identity)}\cr
d = 2: &~~\{\Phi_0,\, \,  \Phi_{K}\}\cr
d = {K + 1\over 2}: &~~\{\Phi_l;~~0\leq {\rm ~even~}l\le K\}{\rm
{}~(the~allowed~integer~spin~representations)}\cr
d = K + 1: &~~\{\Phi_l;~0\leq l\leq K\}\cr}\eqn\clsa$$
Thus in the tensor product theory we know all of the possibilities for
operators of the form $(j,0\vert j,0)$ or $(0,j\vert 0,j)$. Given \nabcon\ and
the uniqueness of the vacuum state $(0,0\vert 0,0)$ in the tensor product
theory, the multiplicities of the operators in the closed sub-algebras must be
as in \clsa.

   We can now write down all of the possibilities for $a,b,K_A$ and $K_B$
consistent with \nabcon\  and \clsa\ , and proceed to consider each
category of possible tensor product modular invariant individually:

\noindent (1) $a = K_A +1, ~~b = K_B+1$ :  \hskip 1.5em
Here we have $N^{AB}_{l\bar l m\bar m} =
\delta_{l\bar l} \delta_{m\bar m} + M^{AB}_{l \bar l m \bar m}$,
with $M^{AB}_{l \bar l m \bar m}$  traceless with respect to both $l, \bar l$
and $m, \bar m$. It is easy to see that $M^{AB}$ must in fact vanish, leaving
us with the simple uncorrelated tensor product of the SU(2)$_{K_A}$ and
SU(2)$_{K_B}$ diagonal modular invariants. Were this not the case, then
$M^{AB}$ by itself would give rise to a modular invariant which did not include
the term containing the identity operator. But this is not possible since
(from \ssu\  and quite generally
in a unitary theory) $S_{0l}>0$ for all $l$.

\noindent (2) $a = {K_A + 1\over 2}, ~~b = {K_B +
1\over2}$ :  \hskip 1.5em
In this case the operators diagonal in either of the factor theories comprise
the set $\{(l,m\vert l,m)\}$ with $l$ and $m$ both odd or both even. This set
contains the operators $(1,K_B\vert 1,K_B)$ and $(K_A,1\vert K_A, 1)$.
The non-diagonal operators in the theory, $(i,j\vert m,l)$ must have a
consistent operator product with these two operators, in particular at least
some of the operators appearing in the naive fusion with them (using the rules
\fra) must have integer spins ($\Delta-\bar{\Delta}\in \Z$). This restricts
the non-diagonal operators $(i,j\vert m,l),~i\ne m,~j\ne l$, to those
satisfying $i+m=K_A$ and $j+l=K_B$. For these operators, in turn, to have
integer spin we have either:  $i-j$ even and $K_A+K_B=0$ modulo 4; or
$i-j$ odd and $K_A-K_B=0$ modulo 4. The former case, taking all such
operators, gives the modular invariant obtained from the
simple tensor product invariant of case (1) by twisting by the simple current
$(K_A,K_B\vert 0,0)$; the latter is obtained by twisting by $(K_A,0\vert
0,K_B)$. An extension of the argument given in (1) using the fact that
$S^{K_A}_{iK_A}S^{K_B}_{jK_B} >0$ for all $i-j$ even, shows that these are the
only possibilities in this category.

\noindent (3) $a=1,~b=2,~K_B=2K_A+1$ or
  $a=2,~b=1,~K_A=2K_B+1$:  \hskip 1.5em
Take $a=1$, $b=2$ so $K_B=2K_A+1$. The model must include
the states $(0,0\vert 0,0)$ and $(0,K_B\vert 0,K_B)$ but no other states of the
form $(0,l\vert 0,l)$, $(i,0\vert i,0)$ or $(j,K_B\vert j,K_B)$.
There must also be two states of
the form $(K_A,j\vert K_A,j)$. Demanding that the fusion products of these
states with themselves are consistent with the above restriction requires
$j=0$, or $K_B$, but then the states themselves are inconsistent with the
restriction. Thus there are no possible consistent theories within this
category.

\noindent (4) $a=2,~b={K_B+1\over 2},~K_A=3$ or
 $a={K_A+1\over 2},~b=2,~K_B=3$: \hskip 1.5em
This case differs from case (2) with $K_A$ and/or $K_B=3$,
in that the $d=2$ closed subalgebra of the SU(2)$_3$ theory consists of
$\{\Phi_0,\Phi_3\}$  instead of $\{\Phi_0,\Phi_2\}$ as in (2). If $a=2$,
$b={K_B+1\over 2}$ and $K_A=3$ then the operators diagonal in either factor
theory comprise the set $\{ (0,l\vert 0,l), (3,l\vert 3,l)~l$ even; $(1,j\vert
1,j), (2,j\vert 2,j)~j$ odd$\}$. There must be additional non-diagonal
operators, $(i,j\vert l,m)$ $i\ne l$, $j\ne m$, if there are to be any modular
invariants in this category. If \hbox {$(i,j\vert l,m)$} appears then
\hbox {$(3-i,j\vert 3-l,m)$}
appears also. For both to have integer spin $i$ and $l$ must be both
even or both odd. Thus there must be operators of the form $(0,j\vert 2,m)$
or $(1,p\vert 3,l)$. Fusing these
with the operators
\hbox {$(1,K_B\vert 1,K_B)$}
from
the diagonal part of the theory produces the operators
\hbox {$(1,K_B-j\vert 1,K_B-m)$}
and/or
\hbox {$(1,K_B-j\vert 3,K_B-m)$} and \hbox {$(0,K_B-p\vert 2,K_B-l)$}
and/or
\hbox {$(2,K_B-p\vert 2,K_B-l)$,}
respectively.
It is easy to see that if the former
fields have integer spin then none of the possible fusion products do. Thus,
there can be no consistent theories in this category.

\noindent (5) $ K_A=K_B\equiv K $, $a= b= 1$ or $a = b = 2$ : \hskip 1.5em
The situation
becomes more complicated for $K_A = K_B \equiv K$.  For these cases we have
additional trace equations,
$$\eqalign{\sum_{l=\bar{m}} N^{AB}_{lm\bar{l}\bar{m}}\, &=a^\prime
\delta_{m\bar{l}}\,\, ;\quad\quad
a^\prime\in \Z^+\cr
 \sum_{m=\bar{l}} N^{AB}_{lm\bar{l}\bar{m}}\, &=b^\prime
\delta_{l\bar{m}}\,\, ;\quad\quad b^\prime\in \Z^+\quad .\cr}\eqn\morcon$$
If the values of $a^\prime$ and $b^\prime$ correspond to any of cases
(1)---(4),
then the invariants are precisely as given above, with the factor
theories permuted. Thus we only need to consider the cases:
(5a) $a=b=a^\prime=b^\prime=1$, (5b) $a=b=a^\prime=b^\prime=2$,
and (5c) $a=b=1,~
a^\prime=b^\prime=2$. Case (5b) is most quickly disposed of. The operators in
the theory include the closed subalgebra $\{(0,0\vert 0,0),(0,K\vert 0,K)\}$,
$\{(0,0\vert 0,0),(K,0\vert K,0)\}$, $\{(0,0\vert 0,0),(K,0\vert 0,K)\},$ and
$\{(0,0\vert 0,0),(0,K\vert K,0)\}$. For the operator algebra with these
together to be closed the chiral fields $(K,K\vert 0,0)$ and $(0,0\vert K,K)$
must appear, but for K odd these do not have integer conformal dimension.
Therefore this case is ruled out.

    Cases (5a) and (5c) can also be ruled out as follows. In both cases there
must be fields $(1,j\vert 1,j)$ and $(p,1\vert p,1)$ with a single choice for
$j$ and $p$ in each case. Consider the four-point correlation function on the
plane $\langle (1,j\vert 1,j)(1,j\vert 1,j)(p,1\vert p,1)(p,1\vert p,1)
\rangle$. In one channel the only possible intermediate state primary fields
which can appear consistent with the restrictions of cases (5a) or (5c) are
$(0,0\vert 0,0)$ and $(2,2\vert 2,2)$. In the cross channels only a subset of
the states of the form $(p\pm 1,j\pm 1\vert p\pm 1,j\pm 1)$ can appear as
intermediates. We know from the four-point amplitudes $\langle (1\vert 1)
(1\vert 1) (j\vert j) (j\vert j)\rangle$ and $\langle (1\vert 1)
(1\vert 1) (p\vert p) (p\vert p)\rangle$ in the factor theories that the chiral
blocks making up the amplitudes have two-dimensional monodromy, so that the
blocks appearing in the tensor product theory have four-dimensional monodromy.
There is no way, then, to assemble the two chiral blocks corresponding to the
allowed intermediate primaries $(0,0\vert 0,0)$ and $(2,2\vert 2,2)$ in such a
way that the four-point function in the tensor product theory can be monodromy
invariant (i.e. single valued).

    To summarize: We have used the constraints \constr\ and the consistency of
conveniently chosen fusion rules and amplitudes to find the only consistent
tensor product theories of the type $SU(2)_{K_A}\otimes SU(2)_{K_B}$, with
$K_A,K_B$ odd. These turn out to be the simple (uncorrelated) product of the
diagonal invariants of the factor theories and all theories obtained from
these ones by twisting by the allowed simple current fields which can be built
from the identity and fields labeled $K_A$ and $K_B$.

\chapter{Left-Right Asymmetric Modular Invariants}

    So far we have considered tensor product conformal field theories  which
are diagonal in the sense that for each holomorphic conformal field theory
factor there is a corresponding anti-holomorphic conformal field theory factor
with the same chiral algebra. While these are the relevant theories to consider
for statistical mechanics applications, it is natural in the construction of
heterotic string theories to consider conformal field theories which are
inherently left-right asymmetric as well. For these the methods discussed above
do not apply. Nonetheless, we can exploit known properties of left-right
symmetric conformal field theories to construct modular invariants even for
inherently asymmetric theories by using the following result: Given two
consistent diagonal rational conformal field theories (apriori with different
chiral algebras) with modular invariant partition functions
$Z^A=\sum\chi^A_i\bar{\chi}^A_i$ and $Z^B=\sum\chi^B_i\bar{\chi}^B_i$,
the left-right asymmetric partition function given by
$Z^{A\bar{B}}=\sum\chi^A_i\bar{\chi}^B_i$ will be modular invariant
 only if: (1) the conformal dimensions
agree modulo 1, or more precisely
$\Delta^A_i-c^A/24=\Delta^B_i-c^B/24\quad ({\rm mod\ 1})$; and (2) the
fusion rules of the two theories coincide, $\phi^A_i\times \phi^A_j=\sum_k
N_{ij}^k\phi^A_k$ and $\phi^B_i\times \phi^B_j=\sum_k N_{ij}^k\phi^B_k$.

   Condition (1) is obviously necessary and sufficient for $Z^{AB}$ to be
$T$ invariant. Condition (2) is almost immediate given Verlinde's
results.$^{[7]}$ $Z^{AB}$ is invariant under the $S$
transformation if and only if the
$S$ matrices implementing the modular transformations on the characters of the
$A$ and $B$ theories coincide, $S^A_{ij}=S^B_{ij}$. As Verlinde showed, the
fusion rule coefficients determine the $S$ matrix,\foot{To be precise,
Verlinde showed that the eigenvalues, $\lambda_i^{(j)}$, of the matrices
$(N_i)_l^{\ k}$ satisfy
$\lambda_i^{(j)}=S_{ij}/S_{0j}$ but there could be an ambiguity in the
 choice of superscript $(j)$ labeling each member of the set
of eigenvalues of $(N_i)_l^{\ k}$. We believe in the present case that this
ambiguity is fixed given $T$ and the requirement $(ST)^3=1$, but have no
proof.} so condition (2) is required
for $S^A=S^B$. In employing this relation it is crucial to define the primary
fields with respect to the full chiral algebra of the theory.

     As a simple example, consider the theories $A=$ SO(31) level 1, and $B=$
E$_8$ level 2, both with central charge $c=31/2$.
The consistent diagonal theories have partition functions, $Z^A=
\chi_0\bar{\chi}_0+ \chi_{{1\over 2}}\bar{\chi}_{{1\over 2}}+
\chi_{{31\over 16}}\bar{\chi}_{{31\over 16}}$ and $Z^B=
\chi_0\bar{\chi}_0+ \chi_{{3\over 2}}\bar{\chi}_{{3\over 2}}+
\chi_{{15\over 16}}\bar{\chi}_{{15\over 16}}$, where the characters are labeled
by the conformal dimension of the associated primary field. The fusion rules in
both theories are like those in the Ising model. The asymmetric partition
function, $Z^{AB}=
\chi^A_0\bar{\chi}^B_0+ \chi^A_{{1\over 2}}\bar{\chi}^B_{{3\over 2}}+
\chi^A_{{31\over 16}}\bar{\chi}^B_{{15\over 16}}$ satisfies conditions (1)
and (2) and so is itself modular invariant. This modular invariant can
also be constructed by choosing appropriate boundary conditions for a
collection of 31 free, real fermions.

    A more interesting example, which cannot be constructed from free bosons or
fermions or by twisting a known invariant by a simple current, is the
following. For the $A$ theory we take the simple tensor product of the diagonal
theories for G$_2$ level 1 and SU(3) level 2; for the $B$ theory the simple
tensor product of the diagonal theories for F$_4$ level 1 and the three state
Potts model. The central charges coincide: $c^A=14/5+16/5=6$; $c^B=26/5+4/5=6$.
The primary fields appearing in each theory are, for G$_2$ level 1
the identity and 7 ($\Delta={2\over 5}$); for SU(3) level 2 the identity, 3 and
$\bar{3}$ ($\Delta={4\over 15}$), 6 and ${\bar 6}$ ($\Delta={2\over 5}$), and 8
($\Delta={3\over 5}$); for F$_4$ level 1 the identity and 26
($\Delta={3\over 5}$); and for
the Potts model the primaries, labeled by their conformal dimensions are 0,
${2\over 5}$, ${2\over 3}$, ${\bar {2\over 3}}$, ${1\over 15}$, and
${\bar {1\over 15}}$.

    To economically list the fusion rules for these theories we can simply list
the non-vanishing three-point amplitudes (where we represent the field by its
conformal dimension). Besides the obvious ones involving the identity operator
($\langle \phi\bar{\phi} 0\rangle$) these are: for G$_2$ level 1
$\langle {2\over 5}, {2\over 5}, {2\over 5}\rangle$; for SU(3) level 2,
$\langle
{4\over 15},{4\over 15},{4\over 15}\rangle$, $\langle {4\over 15},{4\over 15},
\bar{{2\over 3}}\rangle$, $\langle
{4\over 15},\bar{{4\over 15}},{3\over 5}\rangle$, $\langle {4\over 15},{3\over
 5},{2\over 3}\rangle$, $\langle
{2\over 3},{2\over 3},{2\over 3}\rangle$, $\langle {3\over 5},{3\over 5},
{3\over 5}\rangle$, and the conjugates of
these; for F$_4$ level 1, $\langle {3\over 5},{3\over 5},{3\over 5}\rangle$;
 and for the three
state Potts model, $\langle
{1\over 15},{1\over 15},{1\over 15}\rangle$, $\langle {1\over 15},{1\over 15},
\bar{{2\over 3}}\rangle$, $\langle
{1\over 15},\bar{{1\over 15}},{2\over 5}\rangle$, $\langle {1\over 15},{2\over
 5},{2\over 3}\rangle$, $\langle
{2\over 3},{2\over 3},{2\over 3}\rangle$, $\langle {2\over 5},{2\over 5},
{2\over 5}\rangle$, and the conjugates of
these. All of the non-zero fusion rule coefficients, $N_{ijk}$, for these four
theories are equal to one.

    Given the obvious similarities of the fusion rules and conformal dimensions
of these theories, it is not difficult to verify that the asymmetric partition
function given by,
$$\eqalign{Z^{AA^{\prime}BB^{\prime}}=
&\chi^A_{0}\chi^{A^{\prime}}_{0}\bar{\chi}^B_{0}
\bar{\chi}^{B^{\prime}}_{0}+\chi^A_{0}\chi^{A^{\prime}}_{{3\over 5}}
\bar{\chi}^B_{{3\over 5}}
\bar{\chi}^{B^{\prime}}_{0}+\chi^A_{0}\chi^{A^{\prime}}_{{2\over 3}}
\bar{\chi}^B_{0}
\bar{\chi}^{B^{\prime}}_{{2\over 3}}+\chi^A_{0}\chi^{A^{\prime}}_{\bar{{2\over
 3}}}
\bar{\chi}^B_{0}\bar{\chi}^{B^{\prime}}_{\bar{{2\over 3}}}+\cr &\chi^A_{0}
\chi^{A^{\prime}}_{{4\over 15}}\bar{\chi}^B_{{3\over 5}}
\bar{\chi}^{B^{\prime}}_{\bar{{2\over 3}}}+\chi^A_{0}\chi^{A^{\prime}}_{
\bar{{4\over 15}}}
\bar{\chi}^B_{{3\over 5}}\bar{\chi}^{B^{\prime}}_{{2\over 3}}+\chi^A_{{2\over
 5}}
\chi^{A^{\prime}}_{0}\bar{\chi}^B_{0}\bar{\chi}^{B^{\prime}}_{{2\over 5}}+
\chi^A_{{2\over 5}}\chi^{A^{\prime}}_{{3\over 5}}\bar{\chi}^B_{{3\over 5}}
\bar{\chi}^{B^{\prime}}_{{2\over 5}}+\cr &\chi^A_{{2\over 5}}
\chi^{A^{\prime}}_{{2\over 3}}
\bar{\chi}^B_{0}\bar{\chi}^{B^{\prime}}_{\bar{{1\over 15}}}+\chi^A_{{2\over 5}}
\chi^{A^{\prime}}_{\bar{{2\over 3}}}\bar{\chi}^B_{0}\bar{\chi}^{B^{\prime}}_{
{1\over 15}}+
\chi^A_{{2\over 5}}\chi^{A^{\prime}}_{{4\over 15}}\bar{\chi}^B_{{3\over 5}}
\bar{\chi}^{B^{\prime}}_{{1\over 15}}+\chi^A_{{2\over 5}}\chi^{A^{\prime}}_{
\bar{{4\over 15}}}
\bar{\chi}^B_{{3\over 5}}\bar{\chi}^{B^{\prime}}_{\bar{{1\over 15}}}\quad ,\cr}
 $$
satisfies conditions (1) and (2) for modular invariance.
Here $A$, $A^{\prime}$, $B$, and $B^{\prime}$ denote the G$_2$, SU(3), F$_4$,
and Potts theories, respectively. An alternative sewing of the operators in
these four conformal
field theories gives rise to the diagonal E$_6$ level 1 modular invariant,
$$\eqalign{Z^{{\rm E}_6^{(1)}}=
&\chi^A_{0}\chi^{A^{\prime}}_{0}\bar{\chi}^B_{0}
\bar{\chi}^{B^{\prime}}_{0}+\chi^A_{{2\over 5}}\chi^{A^{\prime}}_{{3\over 5}}
\bar{\chi}^B_{0}
\bar{\chi}^{B^{\prime}}_{0}+\chi^A_{0}\chi^{A^{\prime}}_{0}\bar{\chi}^B_{
{3\over 5}}
\bar{\chi}^{B^{\prime}}_{{2\over 5}}+\chi^A_{{2\over 5}}\chi^{A^{\prime}}_{
\bar{{3\over 5}}}
\bar{\chi}^B_{{3\over 5}}\bar{\chi}^{B^{\prime}}_{{2\over 5}}+\cr &\chi^A_{0}
\chi^{A^{\prime}}_{{2\over 3}}\bar{\chi}^B_{0}
\bar{\chi}^{B^{\prime}}_{\bar{{2\over 3}}}+\chi^A_{0}\chi^{A^{\prime}}_{\bar{
{2\over 3}}}
\bar{\chi}^B_{0}\bar{\chi}^{B^{\prime}}_{{2\over 3}}+\chi^A_{0}
\chi^{A^{\prime}}_{{2\over 3}}\bar{\chi^B_{{3\over 5}}}\bar{\chi}^{
B^{\prime}}_{{1\over 15}}+
\chi^A_{0}\chi^{A^{\prime}}_{\bar{{2\over 3}}}\bar{\chi}^B_{{3\over 5}}
\bar{\chi}^{B^{\prime}}_{\bar{{1\over 15}}}+\cr &\chi^A_{{2\over 5}}\chi^{A^{
\prime}}_{{4\over 15}}
\bar{\chi}^B_{{3\over 5}}\bar{\chi}^{B^{\prime}}_{\bar{{1\over 15}}}+\chi^A_{
{2\over 5}}
\chi^{A^{\prime}}_{\bar{{4\over 15}}}\bar{\chi}^B_{{3\over 5}}\bar{\chi}^{B^{
\prime}}_{{1\over 15}}+
\chi^A_{{2\over 5}}\chi^{A^{\prime}}_{{4\over 15}}\bar{\chi}^B_{0}
\bar{\chi}^{B^{\prime}}_{{2\over 3}}+\chi^A_{{2\over 5}}\chi^{A^{\prime}}_{
\bar{{4\over 15}}}
\bar{\chi}^B_{0}\bar{\chi}^{B^{\prime}}_{\bar{{2\over 3}}}\quad .\cr} $$

    It is natural to suppose that the asymmetric modular invariant can be
obtained from the symmetric one by twisting by the appropriate field or fields.
This intuition is correct, but the twisting is not by a simple current
operator, and correspondingly there is no definite algorithm for achieving
it. In the symmetric theory the chiral algebra is enlarged (to
E$_6\otimes$E$_6$). Twisting by a simple current (as considered in the
literature) cannot reduce the chiral algebra, and here gives back the same
theory. There is, however, a candidate field which is primary under the {\it
smaller} chiral algebra of the asymmetric theory and which has simple fusion
rules when defined with respect to this algebra, namely the field $(0,{2\over
3}\vert 0,{2\over 3})$. Twisting $Z^{E_6}$ by this operator, that is throwing
out those operators which fused with $(0,{2\over 3}\vert 0,{2\over 3})$ give
$T$ noninvariant states while adding those $T$ invariant operators which result
from fusing, gives only a subset of the characters in the asymmetric theory. To
get the full set we must add the operators formed by fusing $({2\over
5},{3\over 5}\vert {2\over 5},{3\over 5})$ with itself under the now modified
fusion rules of the new theory (which apriori is an ambiguous procedure).
Similarly, twisting the asymmetric invariant by any combinations of simple
currents in that theory gives back the same model.  In order to obtain
$Z^{E_6}$ we have to twist
by the non-simple current $({2\over 5},{3\over 5}\vert
0,0)$, with suitably modified fusion rules, which again is an ambiguous
procedure.

\chapter{Comments}

   The techniques introduced in section 2 make the classification of modular
invariants for tensor product theories built from a small number of factors at
least feasible. A complete classification of the invariants for
SU(2)$_{K_A}\otimes$SU(2)$_{K_B}$ theories, that is the straightforward
extension of the results of section 2 to even $K$, may, in particular, prove
interesting if there is some generalization of the ADE classification found for
the single theories. Nonetheless, a complete classification for tensor product
theories built with many factors is not practical given the enormous number of
possibilities. For the purposes of string model building a procedure for
constructing any particular invariant, such as that available for free field
constructions, would be advantageous. Perhaps a generalization of the twisting
procedure to operators with nontrivial (or altered) fusion rules, as suggested
by the example in section 3, would prove sufficient. In this regard the results
of [8,9], (which have been extensively exploited recently by Gannon$^{[10]}$)
are intriguing though not yet sufficient. In these works,
 obtaining new tensor product modular invariants
is related to shifting the momentum
lattice of a free boson theory, but at the cost of
sacrificing positivity of the coefficients in the partition function.

   Finally we must stress that the condition of modular invariance alone is
insufficient to guarantee a consistent conformal field theory; for
constructions not based on free fields we must ultimately check that a
consistent operator algebra on the plane exists.

Acknowledgements:  G.C. thanks J.H. Schwarz and K. Li for helpful comments
and suggestions on the preparation of this manuscript.

\def\ul{$\underline{\hbox to .75in{\hfill}}$, }
\noindent {\bf References}\hfill\break
\hbsev {\noindent [1]\hfill} A.~Cappelli, C.~Itzykson, and
J.~Zuber, {\it Nucl. Phys.} {\bf B280 [FS 18]} (1987) 445;\hfill\break
\hbsev {\hfill}  {\it Commun. Math. Phys.} 113(1987) 1.\hfill\break
\hbsev {\noindent [2]\hfill} K.S.~Narain, {\it Phys. Lett.}
B169 (1986) 41. \hfill\break
\hbsev {\noindent [3]\hfill} H.~Kawai, D.~ Lewellen, and S.-H.~Tye,
{\it Nucl. Phys.} {\bf B288} (1987) 1; H.~Kawai,\hfill\break
\hbsev {\hfill} D.~Lewellen, J.A.~Schwartz,
and S.-H.~Tye,
{\it Nucl. Phys.} {\bf B299} (1987) 431.\hfill\break
\hbsev {\noindent [4]\hfill} L.~Dixon, J.~Harvey, C.~Vafa and E.~Witten,
 {\it Nucl. Phys.} {\bf B261} (1985) 651;\hfill\break
\hbsev {\hfill} {\bf B274} (1986) 285.\hfill\break
\hbsev {\noindent [5]\hfill} A.N.~Schellekens and S.~Yankielowicz,
{\it Nucl. Phys.} {\bf B327} (1989) 3;\hfill\break
\hbsev {\hfill} A.N.~Schellekens, {\it Phys. Lett.} 244B (1990) 255;
\hfill\break
\hbsev {\hfill} B.~Gato-Rivera and A.N.~Schellekens,
{\it Nucl. Phys.} {\bf B353} (1991) 519;\hfill\break
\hbsev {\hfill}  CERN-TH.6056/91.\hfill\break
\hbsev {\noindent [6]\hfill} P.~Christe, {\it Phys. Lett.}
188B (1987) 219;
{\it Phys. Lett.} 198B (1987) 215; Ph.D.\hfill\break
\hbsev {\hfill} thesis (1986). \hfill\break
\hbsev {\noindent [7]\hfill} E.~Verlinde, {\it Nucl. Phys.} {\bf B300}
(1988) 360.\hfill\break
\hbsev {\noindent [8]\hfill} N.~Warner, {\it Commu. Math. Phys.}
{\bf 130} (1990) 205.\hfill\break
\hbsev {\noindent [9]\hfill} P.~Roberts and H.~Terao, {\it Int. J. Mod.
Phys.}, {\bf A7} (1992) 2207;\hfill\break
\hbsev {\hfill} P.~Roberts, {\it Phys. Lett.} {\bf B244} (1990)
429.\hfill\break
\hbsev {\noindent [10]\hfill} T.~Gannon, Carleton preprints,
hep-th/9209042,43.\hfill\break\hfill\vfill\end